\documentclass[%
 reprint,
 superscriptaddress,
 amsmath,amssymb,
 aps,
]{revtex4-1}

\usepackage{graphicx}
\usepackage{dcolumn}
\usepackage{bm}
\usepackage{xcolor}
\usepackage{url}

\usepackage{subfigure}
\usepackage{multirow}

\begin{document}


\title{Extraction of the proton mass radius from the vector meson photoproductions near thresholds}

\author{Rong Wang}
\email{rwang@impcas.ac.cn}
\affiliation{Institute of Modern Physics, Chinese Academy of Sciences, Lanzhou 730000, China}
\affiliation{University of Chinese Academy of Sciences, Beijing 100049, China}

\author{Wei Kou}
\email{kouwei@impcas.ac.cn}
\affiliation{Institute of Modern Physics, Chinese Academy of Sciences, Lanzhou 730000, China}
\affiliation{University of Chinese Academy of Sciences, Beijing 100049, China}

\author{Ya-Ping Xie}
\email{xieyaping@impcas.ac.cn}
\affiliation{Institute of Modern Physics, Chinese Academy of Sciences, Lanzhou 730000, China}
\affiliation{University of Chinese Academy of Sciences, Beijing 100049, China}

\author{Xurong Chen}
\email{xchen@impcas.ac.cn}
\affiliation{Institute of Modern Physics, Chinese Academy of Sciences, Lanzhou 730000, China}
\affiliation{University of Chinese Academy of Sciences, Beijing 100049, China}
\affiliation{Guangdong Provincial Key Laboratory of Nuclear Science, Institute of Quantum Matter, South China Normal University, Guangzhou 510006, China}


\date{\today}

\begin{abstract}
We present an analysis of the proton mass radius by studying
the $t$-dependence of the differential cross sections
of the vector meson photoproductions near the thresholds.
At low energy, the photoproduction of a quarkonium off the proton is connected
to the scalar gravitational form factor of the proton,
which is sensitive to the proton mass distribution from the QCD trace anomaly.
Under an assumption of the scalar form factor of dipole form,
the proton mass radius is extracted via
the near-threshold photoproduction data of J/$\psi$, $\phi$ and $\omega$ vector mesons.
The average value of the proton mass radius is estimated
to be $\sqrt{\left<R_{\rm m}^2\right>} = 0.67\pm0.03$ fm,
with the dipole cutoff $m_{\rm s}=1.01\pm0.04$ GeV.
\end{abstract}

\maketitle


In modern physics, the proton is viewed as a confined system
of quarks and gluons with energy and pressure inside
governed by the strong interaction.
The underlying theory of strong interaction is quantum chromodynamics (QCD).
Understanding the color confinement and the mass generation of the proton
in QCD theory is one of the deepest questions in particle and nuclear physics.
As a composite particle, various distributions are often used to describe the form of the proton,
such as the charge distribution, the current distribution and the mass distribution.
Up to date, the charge radius and the magnetic radius of the proton are precisely measured,
with $R_{\rm C}=0.8409\pm0.0004$ fm and $R_{\rm mag.}=0.851\pm0.026$ fm \cite{Zyla:2020zbs}.
However in experiment the mass radius of the proton is still unknown.

In theory, the basic mechanical properties of the proton are well encoded
in the energy-momentum tensor $T_{\mu\nu}$ (EMT) \cite{Kobzarev:1962wt,Pagels:1966zza,Ji:1996nm}.
By investigating the trace of EMT, Ji suggests a decomposition of the proton mass
including the quantum anomalous energy of QCD \cite{Ji:2021pys,Ji:2021mtz,Ji:1994av,Ji:1995sv,Lorce:2017xzd}.
This mass decomposition is further studied with the lattice QCD (LQCD) calculation \cite{Yang:2018nqn}
and the preliminary analysis of J/$\psi$ photoproduction data \cite{Ali:2019lzf,Wang:2019mza}.
Recently the renormalization of EMT and the scheme-dependence
of proton mass decompositions are carefully studied and discussed \cite{Metz:2020vxd,Rodini:2020pis}.
Though there is still a long way to have a final decomposition of the proton mass,
the origin of the proton mass can be understood in QCD theory.

The proton matrix element of EMT contains three gravitational form factors (GFFs)
($A(t)$, $B(t)$, $D(t)$) \cite{Pagels:1966zza,Ji:1996nm,Ji:1997pf,Teryaev:2016edw,Polyakov:2002yz,Polyakov:2018zvc},
which is written as,
\begin{equation}
\begin{split}
\left<p^{\prime}|T_{\mu\nu}|p\right> = \bar{u}^{\prime}
\left[A(t)\frac{\gamma_{\{\mu}P_{\nu\}} }{2}+B(t)\frac{iP_{\{\mu}\sigma_{\nu\}\rho} \Delta^{\rho}}{4m}  \right.\\
\left.+D(t)\frac{\Delta_{\mu}\Delta_{\nu} - g_{\mu\nu}\Delta^2}{4m}
+m\bar{c}(t)g_{\mu\nu}\right]u e^{i\left(p^{\prime}-p\right) x},
\end{split}
\label{eq:GravitationalFF}
\end{equation}
where the kinematic variables are $P=\frac{1}{2}\left(p^{\prime}+p\right)$, $\Delta=p^{\prime}-p$, $t=\Delta^{2}$,
and the covariant normalization is expressed as
$\left\langle p^{\prime} \mid p\right\rangle=2 p^{0}(2 \pi)^{3} \delta^{(3)}(\boldsymbol{p}^{\prime}-\boldsymbol{p})$.
Using the Gordon identity $2m\bar{u}'\gamma^\alpha u=\bar{u}^{\prime}(2 P^{\alpha}+i \sigma^{\alpha \kappa} \Delta_{\kappa})u$
an alternative decomposition is given by \cite{Polyakov:2018zvc},
\begin{equation}
\begin{split}
\left\langle p^{\prime}\left|T_{\mu \nu}\right| p\right\rangle= \bar{u}^{\prime}\left[A(t) \frac{P_{\mu} P_{\nu}}{m}+J(t) \frac{i P_{\{\mu} \sigma_{\nu\} \rho} \Delta^{\rho}}{2 m}\right.\\
\left.+D(t) \frac{\Delta_{\mu} \Delta_{\nu}-g_{\mu \nu} \Delta^{2}}{4 m}+m \bar{c}(t) g_{\mu \nu}\right] u e^{i\left(p^{\prime}-p\right) x}.
\end{split}
\end{equation}
The Fourier transforms of the GFFs $A(t)$, $J(t)$ and $D(t)$
provide the mass distribution \cite{Pagels:1966zza,Ji:1994av,Ji:1995sv,Ji:1996nm,Polyakov:2018zvc},
the angular momentum distribution \cite{Ji:1996nm,Ji:1997pf,Teryaev:2016edw,Ji:1996ek}
and the pressure distribution \cite{Polyakov:2002yz,Polyakov:2018zvc,Burkert:2018bqq} inside the proton respectively.
$A(t)$, $B(t)$ and $D(t)$ are all renormalization scale invariant,
since here the GFFs are defined as the sum of quark and gluon contributions.
As a consequence of momentum conservation, the constraint $A(0)=1$ is given \cite{Ji:1997pf,Teryaev:2016edw,Polyakov:2018zvc}.
Recently the form factor of $T_{00}$ is suggested to be the mass form factor \cite{Ji:2021pys,Ji:2021mtz,Kharzeev:2021qkd}.

On the theoretical side, there are many progresses which have been made on the GFFs,
such as LQCD \cite{Shanahan:2018pib}, holographic QCD \cite{Mamo:2019mka}, light-cone QCD \cite{Azizi:2019ytx},
MIT bag model \cite{Neubelt:2019sou}, Nambu-Jona-Lasinio (NJL) model \cite{Freese:2019bhb},
chiral perturbation theory \cite{Chen:2001pva,Belitsky:2002jp,Diehl:2006js},
instanton model \cite{Polyakov:2018exb}, QCD sum rule \cite{Anikin:2019kwi} and the dispersion relation \cite{Pasquini:2014vua}.
The form factors of the proton are usually described with the dipole form,
which corresponds to an exponential distribution of the concerned physical quantity of the proton.
For the dipole form parametrization $A(t)=A(0)/\left(1-\frac{t}{m_{\rm A}^2}\right)^2$,
LQCD calculation gives $m_{\rm A}=1.13\pm0.06$ GeV \cite{Shanahan:2018pib}
and the soft-wall model of holographic QCD gives $m_{\rm A}=1.124$ GeV \cite{Mamo:2019mka}.
The dipole gravitational form factor also meets the asymptotic behavior of the GFF at large momentum
based on the perturbative QCD \cite{Tong:2021ctu} and the power counting analysis \cite{Brodsky:1973kr,Matveev:1973ra,Ji:2003fw}.

To access the mechanical properties of the proton,
the electromagnetic interaction can be used.
In principle, the GFFs of the proton can be extracted from
the generalized parton distributions (GPD) with
the full $t$-dependence and $\xi$-dependence \cite{Polyakov:2018zvc,Burkert:2018bqq}.
Using the dispersion relation, the GFF $D(t)$ was indirectly extracted from
the deeply virtual Compton scattering (DVCS) data in a recent analysis \cite{Burkert:2018bqq}.
The peak repulsive pressure about $10^{35}$ pascals is deduced from
the analysis of $D(t)$. Moreover, the repulsive pressure and the confining pressure
are balanced around $r=0.6$ fm \cite{Burkert:2018bqq}.
For the pion, the mass radius has been reported to be $0.32\sim 0.39$ fm
by S. Kumano et al, via the measurement of the generalized distribution amplitude (GPA)
of the pion \cite{Kumano:2017lhr}.
The obtained mass radius of the pion is significantly smaller than the charge radius
of the pion, and it is consistent with the NJL model \cite{Freese:2019bhb}.

It is interesting to search for new experimental methods
beside the measurements of GPDs and GPAs to acquire the GFFs of the proton.
In the framework of holographic QCD, the diffractive photoproductions of
heavy quarkonia on the proton are connected to the GFFs of the proton \cite{Hatta:2018ina,Hatta:2019lxo,Boussarie:2020vmu,Mamo:2019mka}.
The near-threshold J/$\psi$ and $\Upsilon$ productions in e-p scattering
and p-A ultraperipheral collision are suggested to study the proton mass problem \cite{Hatta:2019lxo,Boussarie:2020vmu}.
Based on the holographic QCD model,
a recent calculation shows that the elastic proton-proton scattering can be used
to test the GFFs \cite{Xie:2019soz}.
The more reliable way is to employ the van der Waals interaction between the color dipole
and the proton at low energy suggested by D. Kharzeev \cite{Kharzeev:2021qkd,Fujii:1999xn},
which is a famous vector-meson-dominance (VMD) model.
The scalar gravitational form factor of the trace of EMT
is measured via the interaction between the dipole and the proton.
Following Kharzeev's work and assuming a dipole form GFF,
we performed an analysis of the proton mass radius from the differential cross section data
of the near-threshold vector meson photoproductions,
including the data of J/$\psi$, $\phi$ and $\omega$.

In the non-relativistic limit, the mass distribution can be deduced
using the scalar gravitational form factor instead of the form factor of $T_{00}$ \cite{Kharzeev:2021qkd}.
Explicitly, the definition of the mass radius $\left<R_{\rm m}^2\right>$ is given by,
\begin{equation}
\begin{split}
\left<R_{\rm m}^2\right> \equiv \frac{6}{M}\frac{dG(t)}{dt}\big|_{t=0},
\end{split}
\label{eq:MassRadius}
\end{equation}
with $G(0)=M$.
The meaning of the definition of the proton radius is recently discussed
for the measuring processes of both hydrogen spectroscopy and high-energy elastic scattering \cite{Miller:2018ybm}.
The relativistically correct definitions of the proton radius and charge density are introduced
under the perturbation theory of light-front dynamics in the literature \cite{Miller:2018ybm}.
The scalar gravitational form factor is defined as the matrix
element of the trace of EMT, which is written as \cite{Ji:2021pys,Ji:2021mtz,Kharzeev:2021qkd},
\begin{equation}
\begin{split}
G(t)=A(t)M-B(t)\frac{t}{4M}+D(t)\frac{3t}{M}.
\end{split}
\label{eq:ScalarFF}
\end{equation}
The gravitational form factors are usually parameterized as the dipole form.
Therefore we have,
\begin{equation}
\begin{split}
G(t) = \frac{M}{(1-t/m_{\rm s}^2)^2},
\end{split}
\label{eq:DiPoleFF}
\end{equation}
in which $m_{\rm s}$ is a free parameter to be determined in experiment.
According to the definition, the mass radius is connected to
the dipole parameter $m_{\rm s}$ as,
\begin{equation}
\begin{split}
\left<R_{\rm m}^2\right> = \frac{12}{m_{\rm s}^2}.
\end{split}
\label{eq:RadiusAndDipoleParameter}
\end{equation}

The near-threshold cross section of quarkonium photoproduction is directly
related to the matrix element of the scalar gluon operator,
hence the scalar gravitational form factor can be accessed \cite{Kharzeev:2021qkd,Fujii:1999xn}.
The differential cross section of the photoproduction of the quarkonium
can be described with the GFFs,
which is written as
\cite{Kharzeev:2021qkd,Frankfurt:2002ka,Hatta:2018ina,Mamo:2019mka},
\begin{equation}
\begin{split}
\frac{d\sigma}{dt} \propto G^{2}(t).
\end{split}
\label{eq:DiffXsection}
\end{equation}
By studying the differential cross section of the vector meson photoproduction off the proton,
we can extract the parameter $m_{\rm s}$ of the scalar form factor.
Then we acquire the radius information
by using Eq. (\ref{eq:RadiusAndDipoleParameter}).

\begin{figure}[htp]
\centering
\includegraphics[width=0.46\textwidth]{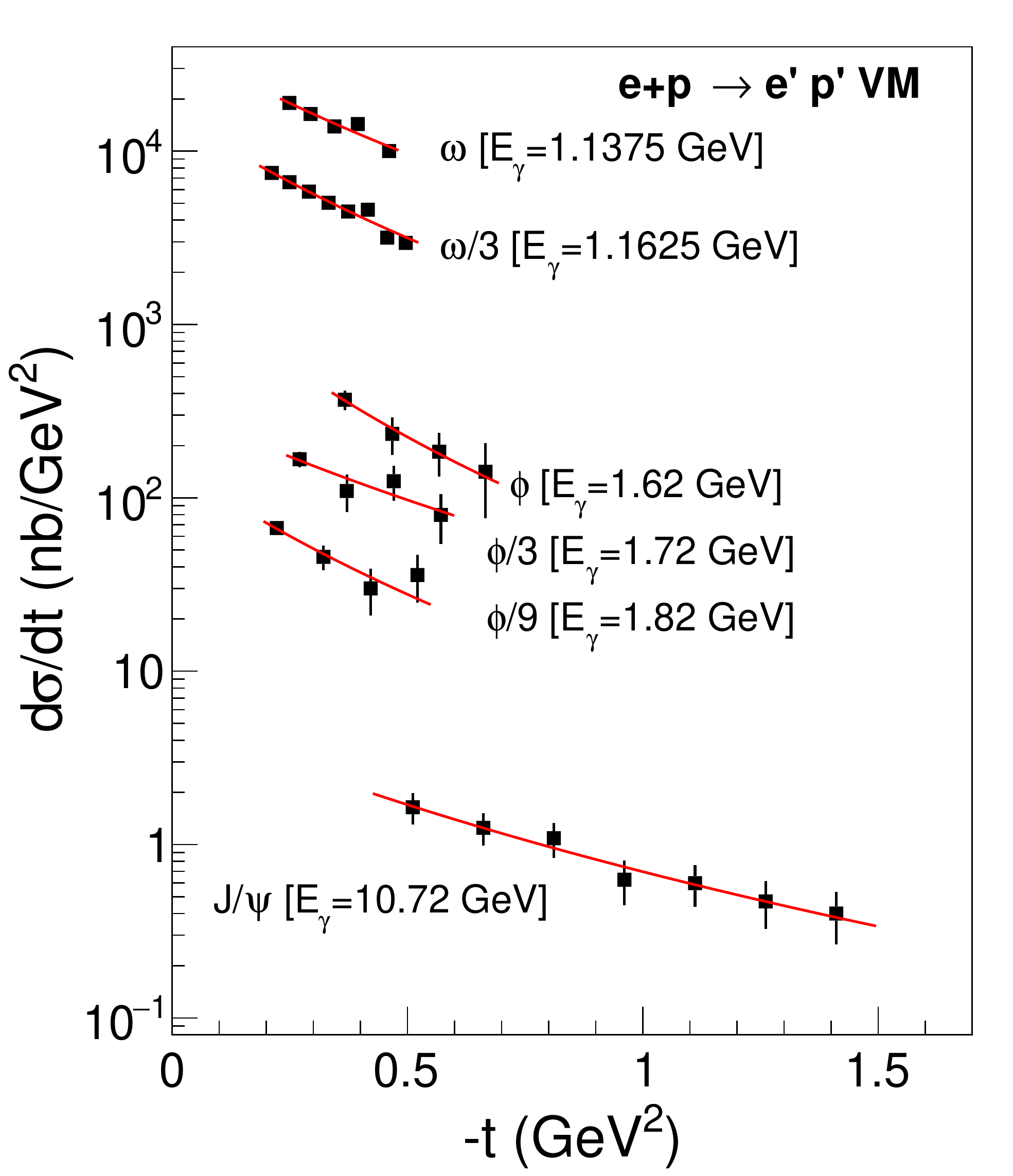}
\caption{
The differential cross sections of the photoproductions
of J/$\psi$ \cite{Ali:2019lzf}, $\phi$ \cite{Mibe:2005er} and $\omega$ \cite{Barth:2003kv}
vector mesons near the thresholds.
The energies of the incident photons are labeled in the figure for the corresponding experimental data.
Some of the cross sections are scaled with the factors,
which are also indicated in the figure.
}
\label{fig:diff-xsection-VMs}
\end{figure}

Fig. \ref{fig:diff-xsection-VMs} shows the differential cross sections of the vector meson
photoproductions as a function of $-t$. The differential cross sections are fitted with
the scalar gravitational form factor of the dipole parametrization (see the curves in Fig. \ref{fig:diff-xsection-VMs}).
The only heavy quarkonium photoproduction data near the threshold is from GlueX  \cite{Ali:2019lzf}.
We determined the dipole parameter to be $m_s=1.23\pm 0.19$ GeV,
and the proton mass radius to be $\sqrt{\left<R_{\rm m}^2\right>}=0.55\pm 0.09$ fm,
from the fit to the J/$\psi$ data.
In experiment, there are already some data on the near-threshold photoproductions of light quarkonia.
We take the differential cross sections of near-threshold $\phi$ photoproduction
at different photon energies by LEPS Collaboration \cite{Mibe:2005er}.
The extracted dipole parameter and the proton mass radii
are listed in Table \ref{tab:RadiusListPhiMeson}, from the fits to the $\phi$ data.
The average mass radius of the three extracted values
is obtained to be $\sqrt{\left<R_{\rm m}^2\right>}=0.67\pm0.10$ fm.
We take the differential cross sections of near-threshold $\omega$ photoproduction
at different photon energies by SAPHIR at ELSA \cite{Barth:2003kv}.
The extracted dipole parameter and the proton mass radii
are listed in Table \ref{tab:RadiusListOmegaMeson}, from the fits to the $\omega$ data.
The average mass radius of the two extracted values
is obtained to be $\sqrt{\left<R_{\rm m}^2\right>}=0.68\pm0.03$ fm.

\begin{figure}[htp]
\centering
\includegraphics[width=0.42\textwidth]{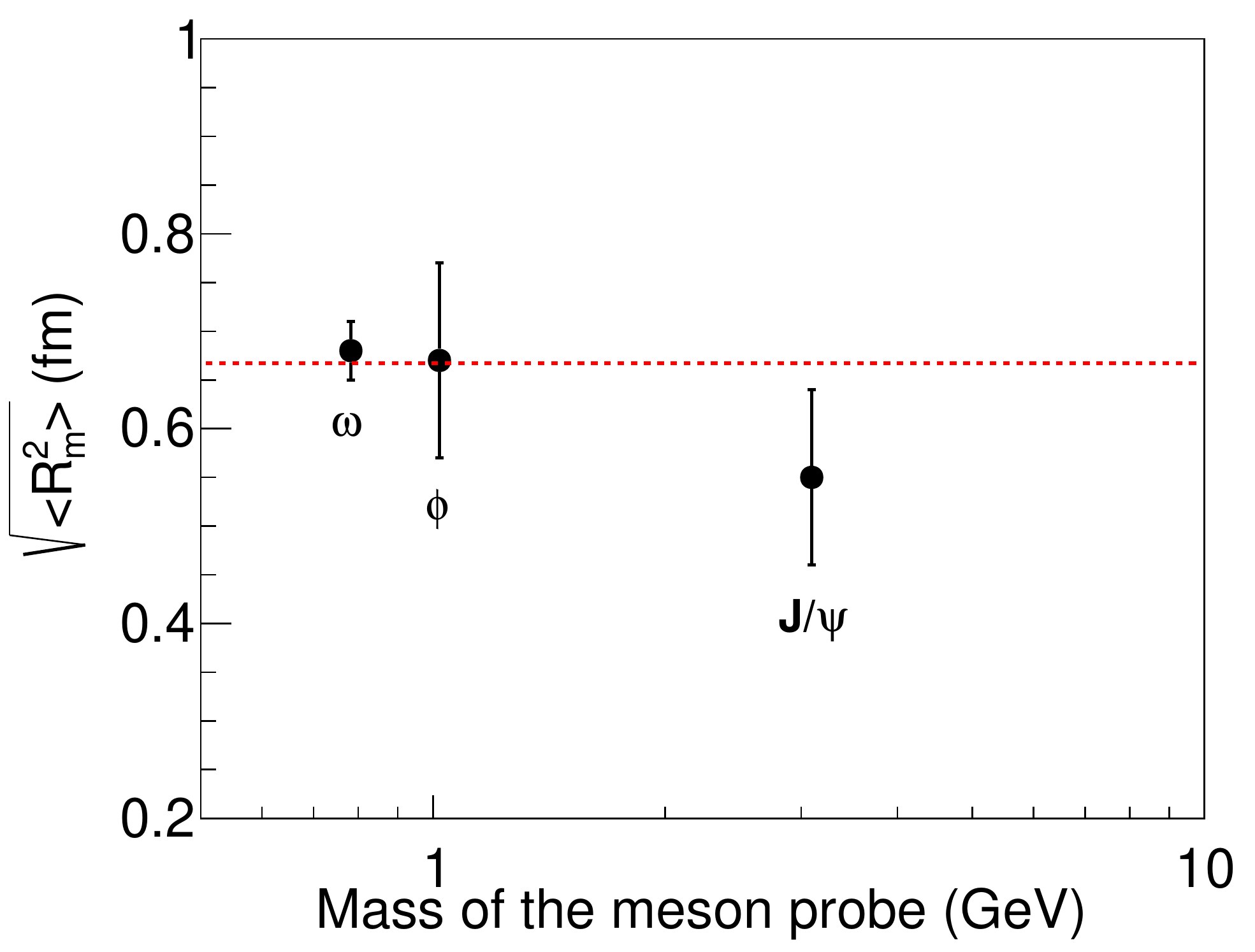}
\caption{
The proton mass radii from different meson production channels.
The dash line shows the average value from the least-square fit.
}
\label{fig:proton-mass-radius}
\end{figure}

Fig. \ref{fig:proton-mass-radius} shows the extracted proton mass radius
as a function of the mass of the vector meson, which comes from
the color dipole interacting with the proton target.
The extracted values are consistent with each other within the statistical uncertainties.
The combined analysis of the data of the three vector mesons gives
the average proton mass radius to be $\sqrt{\left<R_{\rm m}^2\right>} = 0.67\pm0.03$ fm,
with the average dipole parameter to be $m_{\rm s}=1.01\pm0.04$ GeV.

The validity of using Eq. (\ref{eq:DiffXsection}) in describing the light vector meson
photoproduction off the proton is understandable. S. Brodsky et al. suggest that the forward
differential cross section of any possible vector meson leptoproduction can be factorized into
the $q\bar{q}$ wave function of the vector meson and the gluon distribution of the target,
in the region of small momentum transfer \cite{Brodsky:1994kf}. In the nonperturbative approach, phenomenological
Pomeron exchange gives a good description of the diffractive vector meson production under small $t$ \cite{Donnachie:1987pu,Donnachie:1988nj}.
It is commonly agreed that the Pomeron exchange is a good approximation
of the exchange of two gluons. The previous analysis find that the two-gluon form factor
also well describes the $\phi$ photoproduction \cite{Frankfurt:2002ka}.
The $\phi$ and $\omega$ data used in this work are in the small $t$ range ($\sim 0.4$ GeV$^2$),
for which the two-gluon exchange contribution is dominant.
Therefore the formalism in this analysis is approximately effective
in describing the photoproductions of light vector mesons.

\begin{table}[h]
\caption{The extracted values of the dipole parameter and the proton mass radii
from the differential cross sections of $\phi$ photoproduction near threshold
at different photon energies.
}
\begin{center}
\begin{ruledtabular}
\begin{tabular}{ cccc }
  $E_{\gamma}$ (GeV)               &      1.62          &      1.72          &  1.82        \\
  \hline
  $m_{\rm s}$ (GeV)                      &  $0.82\pm 0.24$    &  $1.17\pm 0.30$    &  $0.96\pm 0.20$    \\
  $\sqrt{\left<R_{\rm m}^2\right>}$ (fm)   &  $0.83\pm 0.25$    &  $0.58\pm 0.15$    &  $0.71\pm 0.15$    \\
\end{tabular}
\end{ruledtabular}
\end{center}
\label{tab:RadiusListPhiMeson}
\end{table}

\begin{table}[h]
\caption{The extracted values of the dipole parameter and the proton mass radii
from the differential cross sections of $\omega$ photoproduction near threshold
at different photon energies.
}
\begin{center}
\begin{ruledtabular}
\begin{tabular}{ ccc }
  $E_{\gamma}$ (GeV)               &      1.1375          &      1.1625               \\
  \hline
  $m_{\rm s}$ (GeV)                      &  $1.06\pm 0.06$    &  $0.99\pm 0.04$       \\
  $\sqrt{\left<R_{\rm m}^2\right>}$ (fm)   &  $0.65\pm 0.04$    &  $0.69\pm 0.03$       \\
\end{tabular}
\end{ruledtabular}
\end{center}
\label{tab:RadiusListOmegaMeson}
\end{table}

Now there is a heating discussion on building a low energy
Electron-ion collider in China (EicC), by upgrading the High-Intensity
heavy-ion Accelerator Facility (HIAF) with a polarized electron accelerator \cite{Chen:2018wyz,Chen:2020ijn,Anderle:2021wcy}.
The center-of-mass energy of e-p collision is around 20 GeV,
which provides a good opportunity to study the near-threshold $\Upsilon$ photoproduction
by exploiting the virtual photon flux.
The $\Upsilon$ photoproduction events at EicC are all near the threshold energy
and of the low $Q^2$.
The electroproduction cross section of $\Upsilon$ at EicC energy
has already been estimated using the pomeron exchange model
and the two-gluon exchange model, which is at the magnitude of 100 fb at 20 GeV \cite{Xu:2020uaa}.
According to a recent calculation \cite{Boussarie:2020vmu}, the near-threshold heavy quarkonia electroproductions
at large $Q^2$ provide the important information about the origin
of proton mass and the mass distribution,
which can be realized with the Electron-Ion Collider in the USA (EIC) \cite{Accardi:2012qut}.
The $\Upsilon$ production experiments at EicC and EIC will be a strict test
of the GFFs and the VMD model in the bottom quark sector.
This will improve our understanding on the proton mass distribution.
Excitingly the large intensity facility of the SoLID program at JLab can
provide a precise measurement of the proton mass distribution
with the near-threshold J/$\psi$ electroproduction data.
These future near-threshold photoproductions of heavy quarkonia surely will
play an important role in revealing the mass distribution and the mass radius
of the proton.

The photon-induced vector meson productions off the proton
can be described with the scalar GFF of the dipole form.
By using this model, we have extracted the proton mass radius to be $\sqrt{\left<R_{\rm m}^2\right>} = 0.67\pm0.03$ fm
from a combined analysis of the data of J/$\psi$, $\phi$ and $\omega$,
which is obviously smaller than the charge radius or the magnetic radius of the proton.
This is very similar to the case for the pion \cite{Kumano:2017lhr}.
Our obtained radius is close to the LQCD calculation ($0.62\pm0.03$ fm with $m_{A}=1.13\pm0.06$ GeV) \cite{Shanahan:2018pib}
and the holographic QCD calculation ($0.62$ fm with $m_{A}=1.124$ GeV) \cite{Mamo:2019mka}.
We find that the mass radius equals
the radial distance at somewhere of the proton pressure
$r^2p(r)$ crossing zero \cite{Burkert:2018bqq}.
The light vector meson data show the larger mass radius than that from
the J/$\psi$ data. This is maybe due to the size effect or the mass effect
of the color-dipole, or the validity of applying the VMD model for the near-threshold
light vector meson productions.
The obviously smaller mass radius indicates that the energy distribution is
significantly different from the charge distribution in the proton rest frame.
Since the energy distribution is mainly coupled to the gluons
and the charge distribution is coupled to the quarks,
Kharzeev guesses that the smaller mass radius comes from the smaller gluon radius,
or from the interplay of scale anomaly and spontaneously broken chiral symmetry \cite{Kharzeev:2021qkd}.

\begin{acknowledgments}
We thank Dr. Nu XU for the fruitful discussions.
This work is supported by the Strategic Priority Research Program of Chinese Academy of Sciences
under the Grant NO. XDB34030301
and the National Natural Science Foundation of China under the Grant NO. 12005266.
\end{acknowledgments}

\bibliographystyle{apsrev4-1}
\bibliography{refs}

\end{document}